\documentclass[10pt,twocolumn]{article}

\usepackage[
bookmarks=false
,bookmarksnumbered=true
,hypertexnames=false
,breaklinks=true
]{hyperref}
\hypersetup{
  pdfauthor={Kiseleva et al.},
  pdftitle={Where to Go on Your Next Trip?\\ Optimizing Travel Destinations Based on User Preferences},
  pdfsubject={SIGIR SIRIP'15 submission},
  pdfkeywords={Information Storage and Retrieval},
  pdfcreator={LaTeX with hyperref package},
  pdfproducer={pdflatex}
}

\usepackage{comment}
\usepackage{booktabs}
\usepackage{paralist}
\usepackage{xspace}
\usepackage{graphicx}
\usepackage{graphics}
\usepackage[utf8]{inputenc}
\usepackage{epsfig}
\usepackage{multirow}
\usepackage{amsmath}
\usepackage{mathtools}
\usepackage{algorithm}
\usepackage{algorithmic}
\usepackage{color}

\usepackage[square,comma,numbers,sort&compress,sectionbib]{natbib} 
\def\:{\hskip0pt} 
\usepackage{booktabs}
\usepackage{tabularx}

\title{Where to Go on Your Next Trip?\\ Optimizing Travel Destinations Based on User Preferences}

\author{
\centerline{\begin{tabular}{@{}c@{~~}c@{~~}c@{}} 
\phantom{Mats Stafseng Einarsen\(^3\)}& \phantom{Alexander Tuzhilin\(^1\)}& \phantom{Mats Stafseng Einarsen\(^3\)}
\\[-2em]
Julia Kiseleva\(^1\)& Melanie J.I. Mueller \(^2\) & Lucas Bernardi\(^2\)
\\[0.5ex]
Chad Davis\(^2\)& Ivan Kovacek\(^2\)& Mats Stafseng Einarsen\(^2\)
\\[0.5ex]
Jaap Kamps\(^3\)& Alexander Tuzhilin\(^4\)& Djoerd Hiemstra\(^5\)
\\[0.5ex]
\end{tabular}}\\
\begin{tabular}{c}
\small\(^1\)Eindhoven University of Technology, Eindhoven, The Netherlands\\
\small\(^2\)Booking.com, Amsterdam, The Netherlands\\
\small\(^3\)University of Amsterdam, Amsterdam, The Netherlands\\
\small\(^4\)Stern School of Business, New York University, New York, USA \\
\small\(^5\)University of Twente, Enschede, The Netherlands\\[1ex]
\end{tabular}
}

\newcommand{\rqmain}{How to exploit multi-criteria rating data to rank travel destination recommendations?}

\begin{document}

\maketitle

\begin{abstract}\small
Recommendation based on user preferences is a common task for e-commerce websites. 
New recommendation algorithms are often evaluated by offline comparison to baseline algorithms such as recommending random or the most popular items. Here, we investigate how these algorithms themselves perform and compare to the operational production system in large scale online experiments in a real-world application.
	Specifically, we focus on recommending travel destinations at \url{Booking.com}, a major online travel site, to users searching for their preferred vacation activities. To build ranking models we use multi-criteria rating data provided by previous users after their stay at a destination. We implement three methods and compare them to the current baseline in \url{Booking.com}: random, most popular, and Naive Bayes. 
    Our general conclusion is that, in an online A/B test with live users, our Naive-Bayes based ranker increased user engagement significantly over the current online system.
\end{abstract}

\medskip
\small
\vspace{1mm}
\normalsize

\section{Introduction}
\label{sec:intro}

This paper investigates strategies to recommended travel destinations  for users who provided a list of preferred activities at \url{Booking.com}, a major online travel agent. This is a complex exploratory recommendation task characterized by predicting user preferences with a limited amount of noisy information. In addition, the industrial application setting comes with specific challenges for search and recommendation systems \citep{kohavi_kdd_2014}.  

To motivate our problem set-up, we introduce a service which allows to find travel destinations based on users' preferred activities, called {destination finder.\footnote{\url{http://www.booking.com/destinationfinder.html}}
\begin{figure*}[!tb]
\centerline{%
\includegraphics[width=0.75\textwidth]{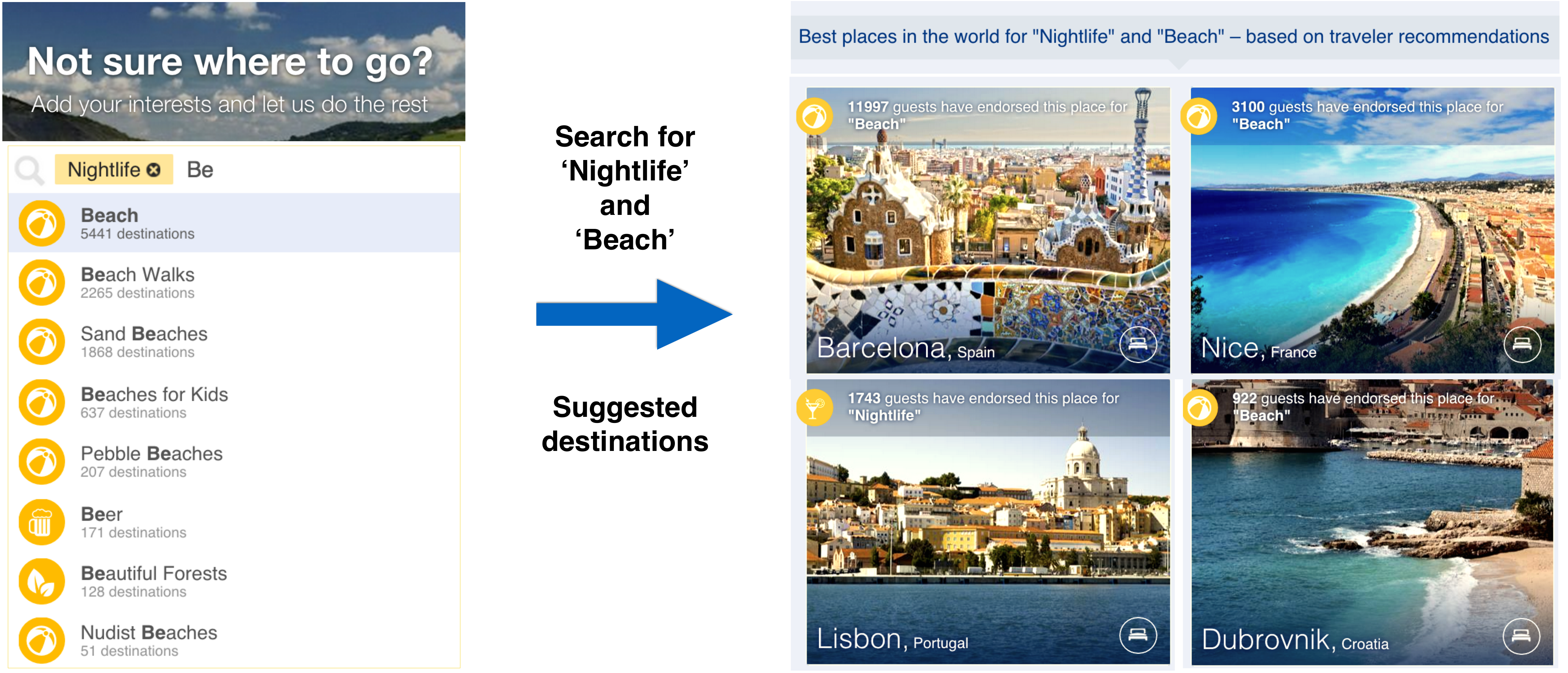}}
\caption{Example of \href{http://www.booking.com/destinationfinder.html}{destination finder} use: a user searching for `Nightlife' and `Beach' obtains a ranked list of recommended destinations (top 4 are shown).}
\label{fig:use_cases}
\end{figure*}
Consider a user who knows what activities she wants to do during her holidays, and is looking for travel destinations matching these activities. This process is a complex exploratory recommendation task in which users start by entering activities in the search box as shown in Figure~\ref{fig:use_cases}. 
The \href{http://www.booking.com/destinationfinder.html}{destination finder} service returns a ranked list of recommended destinations.

The underlying data is based on reviews from users who have booked and stayed at a hotel at some destination in the past.  After their stay, users are asked to endorse the destination with activities from a set of `endorsements'. Initially, the set of endorsements was extracted from users' free-text reviews using a topic-modeling technique such as LDA~\citep{blei_2003, noulas_2014}.  Nowadays, the set of endorsements consists of $256$ activities such as `Beach,'  `Nightlife,' `Shopping,' etc.
These endorsements imply that a user liked a destination for particular characteristics.
Two examples of the collected endorsements for two destinations, `Bangkok' and `London', are shown in Figure~\ref{fig:endors_example}.

As an example of the multi-criteria endorsement data, consider three endorsements: $e_1$ = \emph{`Beach'}, $e_{2}$ = \emph{`Shopping'}, and $e_{3}$ = \emph{`Family Friendly'} and assume that a user $u_j$, after visiting a destination $d_k$ (e.g.\ `London'), provides the review $r_i(u_j,d_k)$ as:
\begin{equation}
\label{eq:review}
r_i(u_j,d_k)=(0,1,0).
\end{equation}
This means our user endorses London for `Shopping' only. However, we cannot conclude that London is not `Family Friendly'. Thus, in contrast to the ratings data in a traditional recommender systems setup, negative user opinions are hidden. In addition, we are dealing with multi-criteria ranking data.

In contrast, in classical formulations of Recommender Systems (RS), the recommendation problem relies on single \emph{ratings} ($R$) as a mechanism of capturing user ($U$) preferences for different items ($I$). The problem of estimating unknown ratings is formalized as follows: $F:U \times I \rightarrow R$.
RS based on latent factor models have been effectively used to understand user interests and predict future actions~\citep{Agarwal_wsdm_2010,Agarwal_kdd_2010}. Such models work by projecting users and items into a lower-dimensional space, thereby grouping similar users and items together, and subsequently computing similarities between them. This approach can run into data sparsity problems, and into a continuous cold start problem  when new items continuously appear.

In multi-criteria RS~\cite{Adomavicius_2007, Adomavicius_2010, Lakiotaki_2011} the rating function has the following form: 
\begin{equation}
\label{eq:mcrs}
F:U \times I \rightarrow (r_0 \times r_1\dots \times r_n)
\end{equation}
The \emph{overall rating} $r_0$ for an item shows how well the user likes this item, while criteria ratings $r_1,\dots,r_n$ provide more insight and explain which aspects of the item she likes.
MCRS predict the overall rating for an item  based on past ratings, using both overall and individual criteria ratings, and recommends to users the item with the best overall score. According to~\cite{Adomavicius_2007}, there are two basic approaches to compute the final rating prediction in the case when the overall rating is known.
In our work we consider a new type of input for RS which is multi-criteria ranking data without an overall rating.

\begin{figure}[!tb]
\centerline{%
\includegraphics[width=\linewidth]{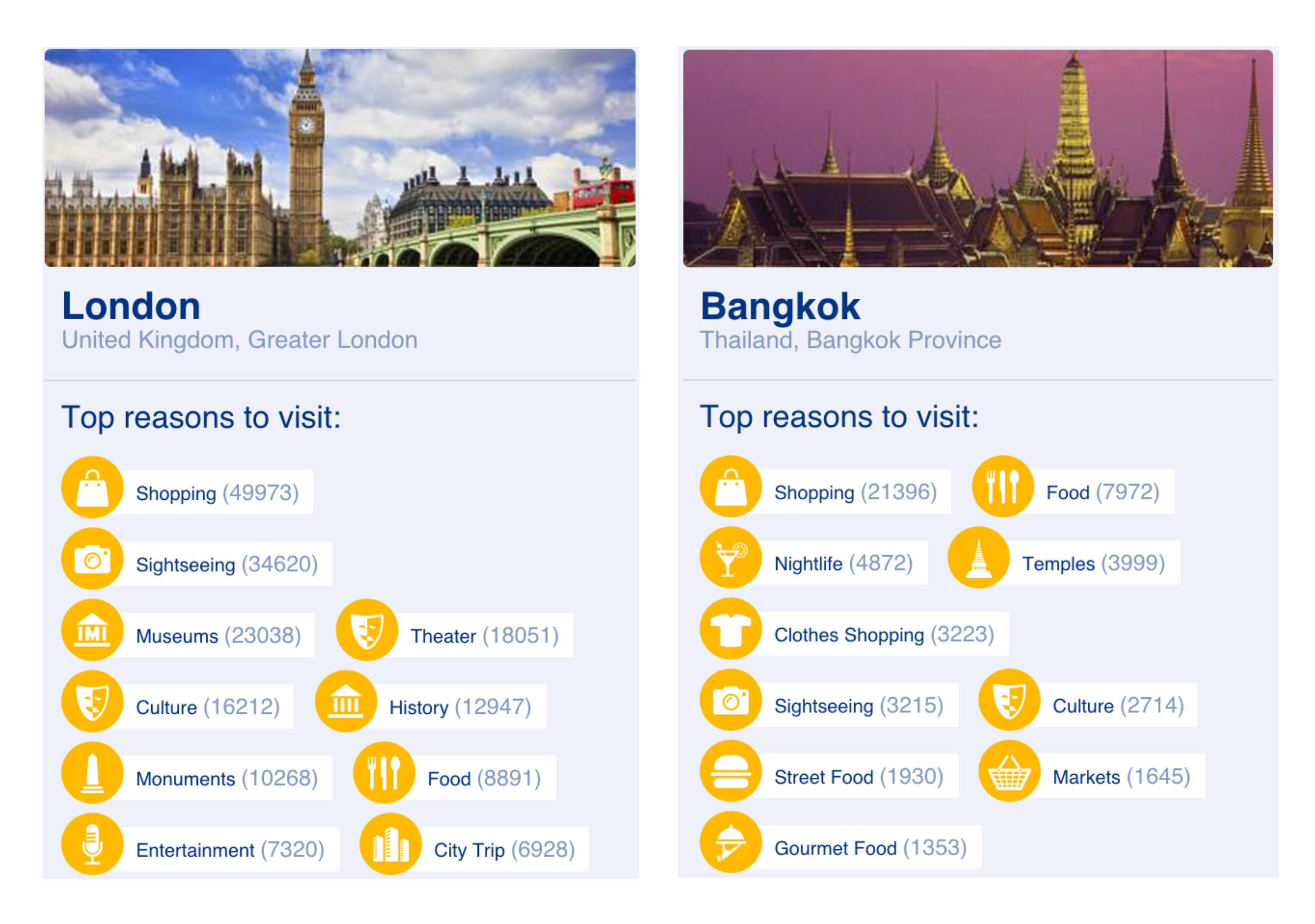}}
\caption{The \href{http://www.booking.com/destinationfinder.html}{destination finder} endorsement pages of \href{http://www.booking.com/destinationfinder/cities/gb/london.html}{London} and \href{http://www.booking.com/destinationfinder/cities/th/bangkok.html}{Bangkok}.}
\label{fig:endors_example}
\end{figure}

There are a number of important challenges in working on the real world application of travel recommendations.

First, it is not easy to apply RS methods in large scale industrial applications. A large scale application of an unsupervised RS is presented in \citep{Hu:2014:SLT:2623330.2623338}, where the authors apply topic modeling techniques to discover user preferences for items in an online store. They apply Locality Sensitive Hashing techniques to overcome performance issues when computing recommendations. We should take into account the fact that if it's not fast it isn't working.
Due to the volume of traffic, offline processing\:---\:done once for all users\:---\:comes at marginal costs, but online processing\:---\:done separately for each user\:---\:can be excessively expensive.   Clearly, response times have to be sub-second, but even doubling the CPU or memory footprint comes at massive costs. 

Second, there is a continuous cold start problem.
A large fraction of users has no prior interactions, making it impossible to use collaborative recommendation, or rely on history for recommendations.   Moreover, for travel sites, even the more active users visit only a few times a year and have volatile needs or different personas (e.g., business and leisure trips), making their personal history a noisy signal at best.

To summarize, our problem setup is the following: 
\textbf{(1)} we have a set geographical destinations such as `Paris', `London', `Amsterdam' etc.; and  
\textbf{(2)} each destination was reviewed by users who visited the destination using a set of endorsements.
Our main goal is to increase user engagement with the travel recommendations as indicator of their interest in the suggested destinations.

Our main research question is: \textsl{\rqmain}
Our main contributions are:
\begin{itemize}
\item we use multi-\:criteria rating data to rank a list of travel destinations;
\item we set up a large-scale online A/B testing evaluation with
live traffic to test our methods;
\item we compared three different rankings against the industrial baseline and obtained a significant gain in user engagement in terms of conversion rates.
\end{itemize}

The remainder of the paper is organized as follows. In Section~\ref{sec:approach}, we introduce our strategies to rank destinations recommendations. We present the results of our large-scale online A/B testing in Section~\ref{sec:experiment}.
Finally, Section~\ref{sec:conclusion} concludes our work in this paper and highlights a few future directions.

\section{Ranking Destination Recommendations}
\label{sec:approach}
In this section, we present our ranking approaches for recommendations of travel destinations. We first discuss our baseline, which is the current production system of the \href{http://www.booking.com/destinationfinder.html}{destination finder} at \url{Booking.com}.  
Then, we discuss our first two approaches, which are relatively straightforward and mainly used for comparison: the random ranking of destinations (Section~\ref{sec:random}), and the list of the most popular destinations (Section~\ref{sec:popularity}).  Finally, we will discuss a Naive Bayes ranking approach to exploit the multi-criteria ranking data.


\subsection{Booking.com Baseline} 

We use the currently live ranking method at \url{Booking.com}'s \href{http://www.booking.com/destinationfinder.html}{destination finder} as a main baseline.  We are not able to disclose the details, but the baseline is an optimized machine learning approach, using the same endorsement data plus some extra features not available to our other approaches. 

We refer further to this method as `Baseline'. 

Next, we present two widely eployed baselines, which we use to give an impression how the baseline performs. Then we introduce an application of the Naive Bayes ranking approach to multi-criteria ranking.

\subsection{Random Destination ranking} 
\label{sec:random}
We retrieve all destinations that are endorsed at least for one of the activities that the user is searching for. The retrieved list of destinations is randomly permuted and is shown to users. 

We refer further to this method as `Random'.

\subsection{Most Popular Destinations} 
\label{sec:popularity}
A very straightforward and at the same time very strong baseline would be the method that shows to users the most popular destinations based on their preferences~\cite{dean2013overview}. For example, if the user searches for the activity `Beach', we calculate the popularity rank score for a destination $d_i$ as the conditional probability: $P(\text{Beach}| d_i)$. 
If the user searches for a second endorsement, e.g.\ `Food', the ranking score for $d_i$ is calculated using a Naive Bayes assumption as: 
$P(\text{Beach}| d_i) \times P(\text{food}| d_i)$. In general, if the users provides $n$ endorsements, $e_1,\ldots,e_n$, the ranking score for $d_i$ is $P(e_1|d_i)\times \ldots \times P(e_n|d_i)$.

We refer further to this method as `Popularity'.

\subsection{Naive Bayes Ranking Approach}
As a primary ranking technique we use a Naive Bayes approach. We will describe its application to the multi-criteria ranking data (presented in Equation~\ref{eq:review}) with an example.
Let us again consider a user searching for `Beach'. We need to return a ranked list of destinations. For instance, the ranking score for the destination `Miami' is calculated as	
\begin{equation}\small
\label{eq:base_ranker}
\begin{split}
P(\text{Miami}, \text{Beach}) =  P (\text{Miami}) \times P(\text{Beach} | \text{Miami}),
\end{split}
\end{equation}
where $P(\text{Beach} | \text{Miami})$ is the probability that  the destination Miami gets the endorsement `Beach'. $P(\text{Miami})$ describes our prior knowledge about Miami. In the simplest case this prior is the ratio of the number of endorsements for Miami to the total number of endorsements in our database. 

If a user uses searches for a second activity, e.g.\ `Food', the ranking score is calculated in the following way:
\begin{equation}\small
\label{eq:base_ranker_2}
\begin{split}
P(\text{Miami}, \text{Beach}, \text{Food}) = 
P (\text{Miami})  \times P(\text{Beach} | \text{Miami}) \\ 
                  \times  P(\text{Food} | \text{Miami})
\end{split}
\end{equation}
If our user provides $n$ endorsements, Equation~\ref{eq:base_ranker_2} becomes a standard Naive Bayes formula.

We refer further to this method as `Naive Bayes'.

\medskip
To summarize, we described three strategies to rank travel destination recommendations: the random ranking, the popularity based ranking, and the Naive Bayes approach. 
These three approaches will be compared to each other and against the industrial baseline.
Next, we will present our experimental pipeline which involves online A/B testing at the \href{http://www.booking.com/destinationfinder.html}{destination finder} service of \url{Booking.com}.

\section{Experiments and Results}
\label{sec:experiment}

In this section we will describe our experimental setup and evaluation approach, and the results of the experiments.  We perform experiments on users of \url{Booking.com} where an instance of the \href{http://www.booking.com/destinationfinder.html}{destination finder} is running in order to conduct an online evaluation. First, we will detail our online evaluation approach and used evaluation measures.  Second, we will detail the experimental results.

\subsection{Research Methodology}

\label{sec:res_meth} 
We take advantage of a production A/B testing environment at \url{Booking.com}, which performs randomized controlled trials for the purpose of inferring causality. A/B testing randomly splits users to see either the baseline or the new variant version of the website, which allows to measure the impact of the new version directly on real users~\citep{koha:onli13,tang_kdd_2010,kohavi_kdd_2014}. 

As our primary \emph{evaluation metric} in the A/B test, we use conversion rate, which is the fraction of sessions which end with at least one clicked result~\citep{lalmas_2014}. 
As explained in the motivation, we are dealing with an exploratory task and therefore aim to increase customer engagement.  An increase in conversion rate is a signal that users click on the suggested destinations and thus interact with the system.  

In order to determine whether a change in conversion rate is a random statistical fluctuation or a statistically significant change, we use the G-test statistic (G-tests of goodness-of-fit). We consider the difference between the baseline and the newly proposed method significant when the G-test p-value is larger than$90\%$.

\subsection{Results}
\label{sec:result}

Conversion rate is the probability for a user to click at least once, which is a common metric for user engagement. We used it as a primary evaluation metric in our experimentation. Table~\ref{tab:res} shows the results of our A/B test. 
The production `Baseline' substantially outperforms the `Random' ranking with respect to conversion rate, and performs slightly (but not significantly) better than the `Popularity' approach. The `Naive Bayes' ranker significantly increases the conversion rate by 4.4\% compared to the production baseline.

\begin{table*}[!tb]
\caption{Results of the \href{http://www.booking.com/destinationfinder.html}{destination finder} online A/B testing based on the number of unique users and clickers. \strut}
\label{tab:res}
\centering
\begin{tabularx}{0.7\linewidth}{X@{~}c@{~}c@{~}c@{~}c}
\toprule
Ranker type & Number of users & Conversion rate & G-test \\
\midrule
Baseline & 9.928 & 25.61\% $\pm$ 0.72\% & \\
Random & 10.079& 24.46\% $\pm$ 0.71\% & 94\% \\
Popularity & 9.838 & 25.50\% $\pm$ 0.73\% & 41\% \\
Naive Bayes & 9.895 & \textbf{26.73\% $\pm$ 0.73\%} & 93\% \\
\bottomrule
\end{tabularx}
\end{table*}

We achieved this substantial increase in conversion rate with a straightforward Naive Bayes ranker.  Moreover, most computations can be done offline. Thus, our model could be trained on large data within reasonable time, and did not negatively impact wallclock and CPU time for the \href{http://www.booking.com/destinationfinder.html}{destination finder} web pages in the online A/B test. This is crucial for a webscale production environment \citep{kohavi_kdd_2014}.

\medskip
To summarize, we used three approaches to rank travel recommendations.  We saw that the random and popularity based ranking of destinations lead to a decrease in user engagement, while the Naive Bayes approach leads to a significant engagement increase.

\section{Conclusion and Discussion}
\label{sec:conclusion}
This paper reports on large-scale experiments with four different approaches to rank travel destination recommendations at \url{Booking.com}, a major online travel agent.  
We focused on a service called \href{http://www.booking.com/destinationfinder.html}{destination finder} where users can search for suitable destination based on preferred activities.  
In order to build ranking models we used multi-criteria rating data in the form of endorsements provided by past users after visiting a booked place.  
    
We implemented three methods to rank travel destinations: Random, Most Popular, and Naive Bayes, and compared them to the current production baseline in \url{Booking.com}.
We observed a significant increase in user engagement for the Naive Bayes ranking approach, as measured by the conversion rate.  
The simplicity of our recommendation models enables us to achieve this engagement without significantly increasing online CPU and memory usage. 
The experiments clearly demonstrate the value of multi-criteria ranking data in a real world application. They also shows that simple algorithmic approaches trained on large data sets can have very good real-life performance \citep{halevy_2009}. 

We are working on a number of extension of the current work, in particular on contextual recommendation approaches that take into account the context of the user and the endorser, and on ways to detect user profiles from implicit contextual information. Initial experiments with contextualized recommendations show that this can lead to significant further improvements of user engagement.

Some of the authors are involved in the organization of the TREC Contextual Suggestion Track \citep{dean2013overview,dean_hall_trec_ca_2014,trec:url}, and the use case of the \href{http://www.booking.com/destinationfinder.html}{destination finder} is part of TREC in 2015, where similar endorsements are collected. The resulting test collection can be used to evaluate destination and venue recommendation approaches.

\medskip
\subsection*{Acknowledgments}
This work was done while the main author was an intern at \url{Booking.com}.
We thank Lukas Vermeer and Athanasios Noulas for fruitful discussions at the early stage of this work.
This research has been partly supported by STW and is part of the CAPA project.\footnote{\url{http://www.win.tue.nl/~mpechen/projects/capa/}}


\renewcommand{\bibsection}{\section*{References}}
\bibliographystyle{abbrvnat}
\setlength{\bibhang}{1em}
\setlength{\bibsep}{.5\itemsep} 

\begin{thebibliography}{15}
\providecommand{\natexlab}[1]{#1}
\providecommand{\url}[1]{\texttt{#1}}
\expandafter\ifx\csname urlstyle\endcsname\relax
  \providecommand{\doi}[1]{doi: #1}\else
  \providecommand{\doi}{doi: \begingroup \urlstyle{rm}\Url}\fi

\bibitem[Adomavicius and Kwon(2007)]{Adomavicius_2007}
G.~Adomavicius and Y.~Kwon.
\newblock New recommendation techniques for multicriteria rating systems.
\newblock \emph{IEEE Intelligent Systems (EXPERT)}, 22\penalty0 (3):\penalty0
  48--55, 2007.

\bibitem[Adomavicius et~al.(2011)Adomavicius, Manouselis, and
  Kwon]{Adomavicius_2010}
G.~Adomavicius, N.~Manouselis, and Y.~Kwon.
\newblock \emph{Multi-Criteria Recommender Systems}, volume 768-803.
\newblock Recommender Systems Handbook, Springer, 2011.

\bibitem[Agarwal and Chen(2009)]{Agarwal_kdd_2010}
D.~Agarwal and B.-C. Chen.
\newblock Regression-based latent factor models.
\newblock In \emph{Proceeding of KDD}, pages 19--28, 2009.

\bibitem[Agarwal and Chen(2010)]{Agarwal_wsdm_2010}
D.~Agarwal and B.-C. Chen.
\newblock flda: matrix factorization through latent dirichlet allocation.
\newblock In \emph{Proceeding of WSDM}, pages 91--100, 2010.

\bibitem[Blei et~al.(2003)Blei, Ng, and Jordan]{blei_2003}
D.~M. Blei, A.~Y. Ng, and M.~I. Jordan.
\newblock Latent dirichlet allocation.
\newblock \emph{Journal of Machine Learning Research}, 3:\penalty0 993--1022,
  2003.

\bibitem[Dean-Hall et~al.(2013)Dean-Hall, Clarke, Kamps, Thomas, Simone, and
  Voorhees]{dean2013overview}
A.~Dean-Hall, C.~L. Clarke, J.~Kamps, P.~Thomas, N.~Simone, and E.~Voorhees.
\newblock Overview of the trec 2013 contextual suggestion track.
\newblock In \emph{Proceeding of TREC}, 2013.

\bibitem[Dean-Hall et~al.(2014)Dean-Hall, Clarke, Kamps, Thomas, and
  Voorhees]{dean_hall_trec_ca_2014}
A.~Dean-Hall, C.~L. Clarke, J.~Kamps, P.~Thomas, and E.~M. Voorhees.
\newblock Overview of the {TREC} 2014 contextual suggestion track.
\newblock In \emph{Proceeding of Text REtrieval Conference (TREC)}, 2014.

\bibitem[Hu et~al.(2014)Hu, Hall, and Attenberg]{Hu:2014:SLT:2623330.2623338}
D.~J. Hu, R.~Hall, and J.~Attenberg.
\newblock Style in the long tail: Discovering unique interests with latent
  variable models in large scale social e-commerce.
\newblock In \emph{KDD}, pages 1640--1649, New York, NY, USA, 2014. ACM.

\bibitem[Kohavi et~al.(2013)Kohavi, Deng, Frasca, Walker, Xu, and
  Pohlmann]{koha:onli13}
R.~Kohavi, A.~Deng, B.~Frasca, T.~Walker, Y.~Xu, and N.~Pohlmann.
\newblock Online controlled experiments at large scale.
\newblock In \emph{Proceedings of KDD}, pages 1168--1176, 2013.

\bibitem[Kohavi et~al.(2014)Kohavi, Deng, Longbotham, and Xu]{kohavi_kdd_2014}
R.~Kohavi, A.~Deng, R.~Longbotham, and Y.~Xu.
\newblock Seven rules of thumb for web site experimenters.
\newblock In \emph{Proceeding of KDD}, pages 1857--1866, 2014.

\bibitem[Lakiotaki et~al.(2011)Lakiotaki, Matsatsinis, and
  Tsoukias]{Lakiotaki_2011}
K.~Lakiotaki, N.~F. Matsatsinis, and A.~Tsoukias.
\newblock Multicriteria user modeling in recommender systems.
\newblock \emph{IEEE Intelligent System}, 26\penalty0 (2):\penalty0 64--76,
  2011.

\bibitem[Lalmas et~al.(2014)Lalmas, O'Brien, and Yom-Tov]{lalmas_2014}
M.~Lalmas, H.~O'Brien, and E.~Yom-Tov.
\newblock Measuring user engagement.
\newblock \emph{Synthesis Lectures on Information Concepts, Retrieval, and
  Services}, 6\penalty0 (4):\penalty0 1--132, 2014.

\bibitem[Noulas and Einarsen(2014)]{noulas_2014}
A.~Noulas and M.~S. Einarsen.
\newblock User engagement through topic modelling in travel.
\newblock In \emph{Proceeding of the Second Workshop on User Engagement
  Optimization}, 2014.

\bibitem[Tang et~al.(2010)Tang, Agarwal, O'Brien, and Meyer]{tang_kdd_2010}
D.~Tang, A.~Agarwal, D.~O'Brien, and M.~Meyer.
\newblock Overlapping experiment infrastructure: More, better, faster
  experimentation.
\newblock In \emph{Proceedings of KDD}, pages 17--26, Washington, DC, 2010.

\bibitem[TREC()]{trec:url}
TREC.
\newblock Contextual suggestion track.
\newblock Text REtrieval Conference, 2015.
\newblock URL \url{https://sites.google.com/site/treccontext/}.

\end{thebibliography}

\end{document}